\documentclass[lettersize,journal]{IEEEtran}
\usepackage{cite}
\usepackage{amsmath,amssymb,amsfonts}
\usepackage{graphicx}
\usepackage{textcomp}
\usepackage{xcolor}
\usepackage{booktabs}
\usepackage{multirow}
\usepackage{url}
\usepackage[colorlinks=false]{hyperref}
\def\BibTeX{{\rm B\kern-.05em{\sc i\kern-.025em b}\kern-.08em
		T\kern-.1667em\lower.7ex\hbox{E}\kern-.125emX}}
\usepackage{balance}
\usepackage{setspace}

\begin{document}
\title{Joint Outage Detection and Compensation for Self-Healing 5G RAN via Deep Reinforcement Learning}

\author{Sajjad~Hussain
\thanks{Manuscript submitted June 26th, 2026.
Sajjad Hussain is with the National University of Sciences and Technology (NUST), Pakistan. Email: sajjad.hussain2@seecs.edu.pk}}

\markboth{Submitted to IEEE WCL, June 2026}%
{Shell \MakeLowercase{\textit{et al.}}: A Sample Article Using IEEEtran.cls for IEEE Journals}

\maketitle

\begin{abstract}
	Self-healing radio access network (RAN) requires autonomous detection and compensation of base station (BS) failures. This letter proposes an end-to-end framework combining three-class cell outage detection (COD), distinguishing	normal, failed, and collaterally degraded cells, with
	a deep Q-Network (DQN) based deep reinforcement learning (DRL) agent that jointly	controls power and antenna tilt for cell outage compensation (COC). Evaluation results show that the proposed DQN agent achieves 99.1\% coverage
	and 54\% full-recovery rate, an 11$\times$ improvement	over the best heuristic, while consuming less compensation	energy than heuristic baselines and learning,
	without explicit geometric input, to prefer tilt-only compensation for centre-cell outage.
\end{abstract}


\section{Introduction}
\label{sec:intro}

\IEEEPARstart{D}{ense} 5G deployments increase BS failure probability, placing tens of user equipment (UEs) in outage and degrading quality of service across coverage zones. The 3GPP Self-Organising Network (SON) specification mandates autonomous self-healing comprising COD and COC~\cite{3gpp_tr36902}. Several DRL-based approaches address coverage, capacity, and resilience optimisation in this context. Hasan and Khalid~\cite{hasan2026} used multi-agent DRL for autonomous per-BS downtilt control, improving edge spectral efficiency by up to 41\%. Kaada \emph{et al.}~\cite{kaada2024} jointly optimized antenna tilt and transmit power using multi-agent DRL, achieving 50--60\% throughput gains and up to 99\% coverage. Liu \emph{et al.}~\cite{liu2024} applied transformer- and CNN-based self-supervised multi-agent DRL to large-scale coverage optimization, reporting coverage improvements of up to 95\%. Raza \emph{et al.}~\cite{raza2025} combined XGBoost-based COD with a DRL agent that jointly optimizes antenna tilt and transmit power, later extending the framework to multi-agent compensation for multiple simultaneous BS failures. Onireti \emph{et al.}~\cite{onireti2015} proposed a $k$-nearest-neighbour COD model integrated with an actor--critic DRL compensation agent. However, both approaches introduce significant computational complexity and may be challenging to deploy within existing network infrastructures.

The contributions of this letter are:
\begin{itemize}
	\item An integrated three-class COD with DRL-COC pipeline: a classifier distinguishing root-cause outage from collaterally degraded	neighbours feeds directly into the DRL agent's	observation state.
	
	\item A joint power-and-tilt action space, with	the agent shown to learn a geometry-aware policy achieving superior coverage, full-network restoration, and compensation energy efficiency than heuristic baselines. \footnote{The codebase is publicly available at: \url{https://github.com/sajjadhussa1n/self-healing-ran}}
\end{itemize}

\section{System Model}
\label{sec:system}

\begin{figure}[!t]
	\centering
	\includegraphics[width=\columnwidth]
	{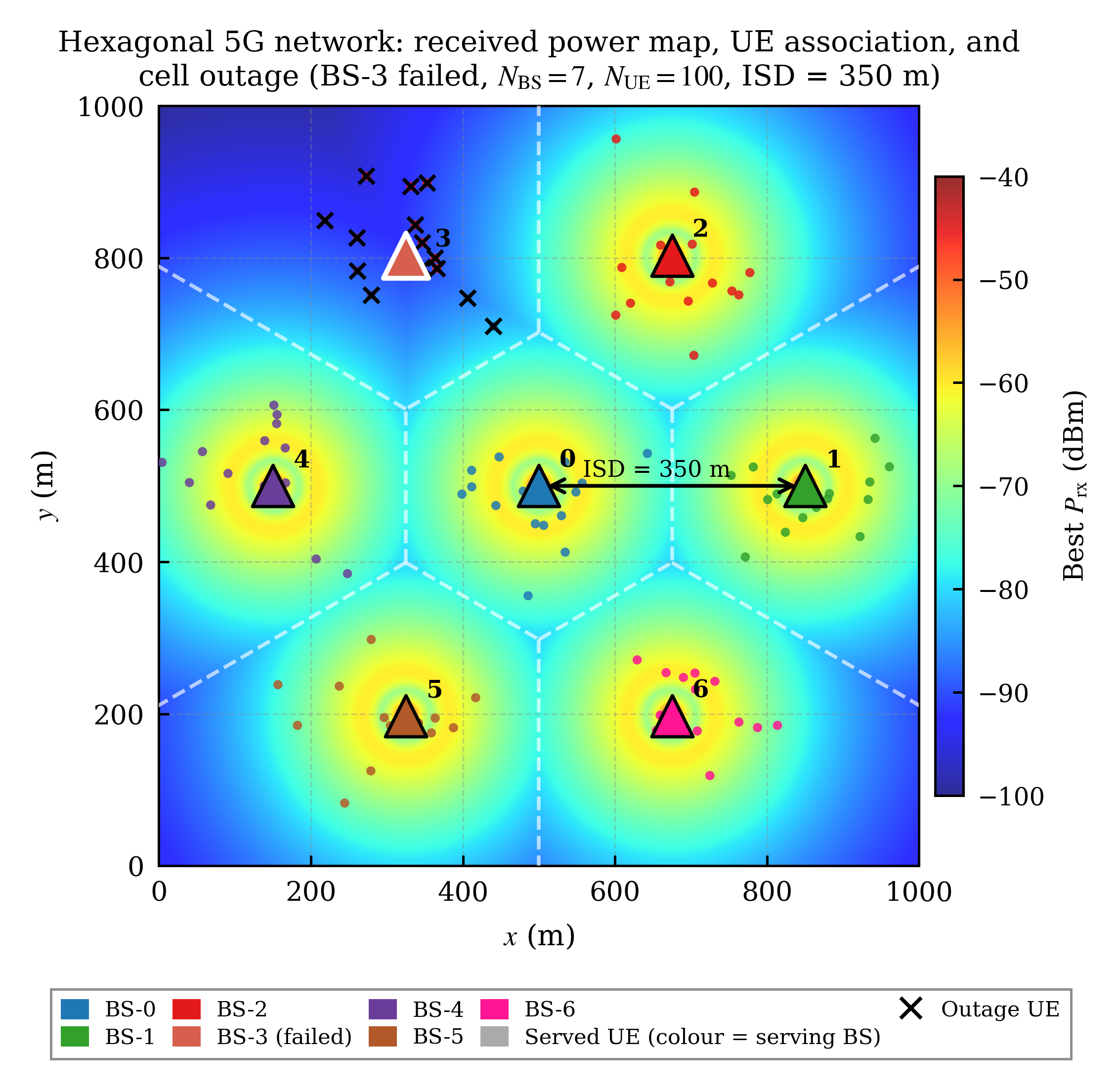}
	\caption{Seven-BS hexagonal network
		($\rm ISD = 350$~m) with Gaussian UE clusters
		(coloured dots) and Voronoi boundaries.
		BS-3 shown failed (red~{\texttimes}).}
	\label{fig:system}
\end{figure}

We consider a downlink single-tier network comprising $N_{\rm BS} = 7$ macro base stations, each serving a single cell, arranged in a hexagonal layout with one BS at the centre and six at the vertices.as illustrated in Fig.~\ref{fig:system}. $N_{\rm UE} = 100$ UEs are distributed using a Gaussian cluster model around each BS, with mobility
modelled as a random walk. Key deployment parameters follow the 3GPP Urban Macro (UMa) scenario~\cite{3gpp_tr36942} and are summarised in Table~\ref{tab:params}.

The received power at UE $k$ from BS $b$ is
\begin{equation}
P_{\rm rx}(k,b) = P_{\rm tx}(b)
- \mathrm{PL}(d_{kb})
+ G_A(\phi_{kb}, \theta_b),
\label{eq:rx}
\end{equation}
where $d_{kb}$ is the 2D distance between UE $k$ and BS $b$, and $G_A(\phi_{kb}, \theta_b)$ is the antenna gain defined in~\eqref{eq:ant}. Path loss follows the log-distance model
\begin{equation}
\mathrm{PL}(d) = 20\log_{10}\!\left(
\frac{4\pi f_c}{c}\right)
+ 10\,n\,\log_{10}(d),
\label{eq:pl}
\end{equation}
with path loss exponent $n = 3.5$. Each UE associates to the BS offering the highest received power, and the resulting SINR at UE $k$ served by BS $b^*$ is
\begin{equation}
\gamma_k =
\frac{P_{\rm rx}(k, b^*)}
{\displaystyle\sum_{b \neq b^*}
	P_{\rm rx}(k, b) + \sigma_n^2},
\label{eq:sinr}
\end{equation}
where $\sigma_n^2$ is the thermal noise power. A UE is declared in outage if $\gamma_k$ falls below a threshold $\gamma_{\rm th}$ or its received power falls below a minimum floor, representing conditions under which reliable decoding is no longer feasible.

The vertical radiation pattern of each BS antenna follows the 3GPP TR~36.942 model~\cite{3gpp_tr36942},
\begin{equation}
G_A(\phi, \theta_b) =
-\min\!\left(
12\left(\frac{\phi - \theta_b}
{\theta_{\rm 3dB}}\right)^{\!2},\;
\mathrm{SLA}
\right) \;\text{[dB]},
\label{eq:ant}
\end{equation}
where $\phi = \arctan\!\bigl(
(h_{\rm BS} - h_{\rm UE})/d\bigr)$
is the vertical angle from the BS to the UE and $\theta_b$ is the electrical downtilt.

A complete BS failure is modelled as $P_{\rm tx}(b) \to -\infty$, representing hardware faults or crashes in which the BS ceases transmission while remaining visible to the
network management system (the \textit{sleeping	cell} problem); outaged UEs reassociate to the next strongest BS at reduced SINR. The network geometry gives rise to two
structurally distinct scenarios: in \textit{edge-BS outage}, asymmetric geometry leaves two to three close neighbours available, making targeted power boosting effective, whereas in \textit{centre-BS outage}, all six neighbours are equidistant, creating a \textit{symmetric interference trap} in which boosting all neighbours simultaneously raises signal and interference equally, yielding negligible SINR improvement. This distinction motivates the joint power-and-tilt action space in Section~\ref{sec:coc}.

\section{Three-Class Cell Outage Detection}
\label{sec:cod}

Prior COD work treats outage detection as binary classification, missing an operationally important third state: a neighbouring BS that absorbs outaged UEs after a nearby failure experiences elevated load and degraded SINR is neither failed nor normal. We propose a three-class scheme: \textit{Normal} (0); \textit{Outage} (1) for the root-cause failed BS; and \textit{Degraded} (2) for active neighbours whose mean SINR drops more than 3~dB from baseline or whose outage UE count exceeds two.
The degraded label identifies collateral victims for targeted compensation, giving the DRL agent a richer state representation than binary COD.

The feature vector comprises own-cell and neighbour-cell KPIs: served UE count $N_b$; PRB load $\rho_b{=}\min(N_b/\bar{N},1)$, where $\bar{N}$ is the network-average UE count per cell, a practical utilisation proxy when scheduling traces are unavailable, capped at unity to flag absorption of a neighbour's UEs; the UE-count ratio
$r_b{=}N_b/N_b^{\rm prev}$; and deltas $\Delta N_b,\Delta\rho_b$. These are mirrored for the 3 nearest BSs, yielding $\mathbf{f}\in\mathbb{R}^{17}$.

The labelled dataset comes from 300 simulated episodes (15 normal $+$ 10 post-outage timesteps each), giving $\approx$52,500 KPI snapshots, with an 80/20 episode-level train/test split to prevent temporal leakage. Table~\ref{tab:cod} compares a threshold rule we define, Logistic Regression, and a 100-tree Random Forest.
The rule flags Outage when $N_b$ drops by $\geq$80\% of its previous value \emph{and} a neighbour's UE count rises by $\geq$2 (transition), or when $N_b{\leq}1$
\emph{and} a neighbour shows an elevated or rising UE count (sustained state); the latter addresses the limitation of pure delta-based rules, which fire only
at the drop's onset timestep. Degraded is flagged when a BS's own UE count rises by $\geq$2 while a neighbour's falls by $\geq$2, i.e., direct evidence of outaged UE absorption.

\begin{table}[!t]
	\caption{Three-Class COD Performance (F1-Score)}
	\label{tab:cod}
	\centering
	\small
	\setlength{\tabcolsep}{6pt}
	\begin{tabular}{lccc}
		\toprule
		\textbf{Classifier}
		& \textbf{Normal}
		& \textbf{Outage}
		& \textbf{Degraded} \\
		\midrule
		Threshold Rule
		& 0.960 & 1.000 & 0.106 \\
		Logistic Regression    & 0.954 & 1.000 & 0.649 \\
		\textbf{Random Forest} & \textbf{0.955}
		& \textbf{1.000}
		& \textbf{0.651} \\
		\bottomrule
	\end{tabular}
\end{table}

All detectors achieve perfect F1-score on label~1, confirming that complete BS failure produces an unambiguous signature. The degraded class is the principal challenge: unlike the Outage branch, the rule's Degraded condition has no sustained-state fallback, firing only at the timestep of the UE-count shift and missing the persistently
degraded state thereafter (F1-score = $0.106$). Random Forest resolves this by learning the sustained feature levels characterising label~2 (F1-score = $0.651$), motivating future LSTM-based temporal classification.

\section{DRL-Based Cell Outage Compensation}
\label{sec:coc}

Given a detected outage at BS $b^* \in \mathcal{B}$, the COC problem is to select a sequence of compensation actions maximising network coverage while minimising collateral degradation. Formally, we seek a policy $\pi^*$ solving
\begin{equation}
\pi^* = \arg\max_{\pi}
\;\mathbb{E}_{\pi}\!\left[
\sum_{t=0}^{T} \gamma^t\, r_t
\;\middle|\; \mathbf{s}_0,\, b^*
\right],
\label{eq:opt}
\end{equation}
where $r_t$ penalises any UE that was covered prior to the failure and is newly pushed into outage by the compensation action at step $t$, as detailed in the
reward formulation below.  We model this as a finite-horizon Markov Decision Process MDP $(\mathcal{S}, \mathcal{A}, \mathcal{R}, \mathcal{P}, \gamma{=}0.95)$ with episode horizon $T = \max(5,\min(\lfloor 0.8\,N_{\rm out} \rfloor + 3,\,10))$ scaling with outage severity $N_{\rm out}$ ($N_{\rm out}$ is the number of UEs already in outage
at episode onset due to the failure of BS), ensuring sufficient steps for large outages without wasteful exploration in simple ones.

The state vector $\mathbf{s}_t \in \mathbb{R}^{52}$ comprises six normalized features for each BS: UE count $\tilde{N}_b \in [0,2]$, mean SINR $\tilde{\gamma}_b \in [0,1]$, signed UE-count variation $\Delta\tilde{N}_b \in [-1,1]$, antenna tilt $\tilde{\theta}_b \in [0,1]$, and transmit power $\tilde{P}_b \in [-1,1]$. These per-BS features are augmented with 10 global features, including the failed-BS one-hot indicator provided by the COD module, the network coverage fraction $\eta_t$, and the number of outage UEs. Values of $\tilde{N}_b>1$ indicate BS overload caused by outaged UEs, while the power sentinel value $\tilde{P}_b=-1$ uniquely identifies the failed BS without requiring an additional binary flag.

The action space $\mathcal{A} = \{A_0,\ldots,A_6\}$ comprises the null action $A_0$ (no compensation) and six strategies of increasing sophistication, denoted S1--S6 in the discussion that follows. S1 applies a fixed power boost to all neighbouring BSs; S2 scales the boost proportionally to each neighbour's load share; S3 concentrates the full power budget on the single neighbour best positioned to rescue the most outaged UEs; S4 distributes power simultaneously across all neighbours weighted by their individual rescue contribution; S5 searches over candidate downtilt reductions, applying the smallest tilt change that maximises rescued UEs subject to minimum collateral
outage; and S6 jointly optimises tilt and power, applying the safe tilt from S5 and adding power only if it rescues additional UEs without violating the collateral constraint.

The reward function
\begin{equation}
r_t = \underbrace{2\,N_{\rm res}^{(t)}}_{\text{rescue}}
- \underbrace{1.5\,N_{\rm coll}^{(t)}}_{\text{collateral}}
+ \underbrace{5\,\Delta\eta_t}_{\text{coverage}}
+ \underbrace{(\eta_t - \eta_0)
	\cdot\mathbb{1}[\eta_t > \eta_0]}_{\text{maintenance}}
\label{eq:reward}
\end{equation}
comprises four terms. The \emph{rescue} term rewards $+2$ per UE $N_{\rm res}^{(t)}$ restored from outage at step $t$, and the \emph{collateral} term penalises $-1.5$ per
previously-covered UE $N_{\rm coll}^{(t)}$ newly pushed into outage; The \emph{coverage} term rewards the per-step change $\Delta\eta_t{=}\eta_t-\eta_{t-1}$ in coverage ratio, providing a dense gradient signal even on steps that do not fully resolve an outage. The \emph{maintenance} term, $(\eta_t-\eta_0)\cdot \mathbb{1}[\eta_t{>}\eta_0]$, rewards the agent for \emph{holding} coverage above the post-outage baseline $\eta_0$ once gains have been made, preventing the policy from undoing earlier rescues or oscillating between actions once a good state is reached; it is zero whenever $\eta_t\leq\eta_0$, so it never rewards inaction before any improvement has occurred. Terminal bonuses of $+10$ and $-5$ are awarded on full restoration and timeout, respectively, and $A_0$ yields $r_t\approx0$ in the absence of prior rescues, ensuring genuinely improving actions are always preferred over inaction.

A DQN agent with MLP policy and network architecture $[256{\to}256{\to}128{\to}7]$ is trained for 50,000 environment steps ($\approx$9200 episodes), taking $\approx$84~min on CPU with $\varepsilon$-greedy exploration decaying from 1.0 to 0.05 over the first 50\% of training. A curriculum learning approach restricts outages to edge BSs for the
first 2,500 episodes before introducing the centre-BS case. A PPO agent with identical architecture is also trained and serves as a comparison baseline. The key
parameters are summarised in Table~\ref{tab:params}.

\begin{table}[!t]
	\caption{System, Channel, and DRL Parameters}
	\label{tab:params}
	\centering
	\resizebox{\columnwidth}{!}{%
	\begin{tabular}{lc|lc}
		\toprule
		\textbf{Parameter} & \textbf{Value}
		& \textbf{Parameter} & \textbf{Value} \\
		\midrule
		Num. BSs ($N_{\rm BS}$)         & 7
		& Num. States ($|\mathbf{s}|$)              & 52            \\
		Inter-site dist. (ISD)                  & 350~m
		& Num. Actions ($|\mathcal{A}|$)             & 7             \\
		Num. UEs ($N_{\rm UE}$)         & 100
		& Discount Factor $\gamma$                    & 0.95          \\
		UE cluster std       & 60~m
		& Learning rate $\alpha$                    & $3\!\times\!10^{-4}$ \\
		TX power             & 43~dBm
		& Network                   & [256,256,128] \\
		Path loss exp.\,$n$  & 3.5
		& Training steps               & 50{,}000      \\
		Noise power          & $-104$~dBm
		& Curriculum eps.              & 2{,}500       \\
		$\gamma_{\rm th}$    & 0~dB
		& Max.\ boost (edge)          & 6~dB          \\
		Tilt range           & $6^\circ$--$25^\circ$
		& Max.\ boost (centre)        & 14~dB         \\
		\bottomrule
	\end{tabular}}
\end{table}

\section{Results}
\label{sec:results}

\begin{figure}[!t]
	\centering
	\includegraphics[width=\linewidth]{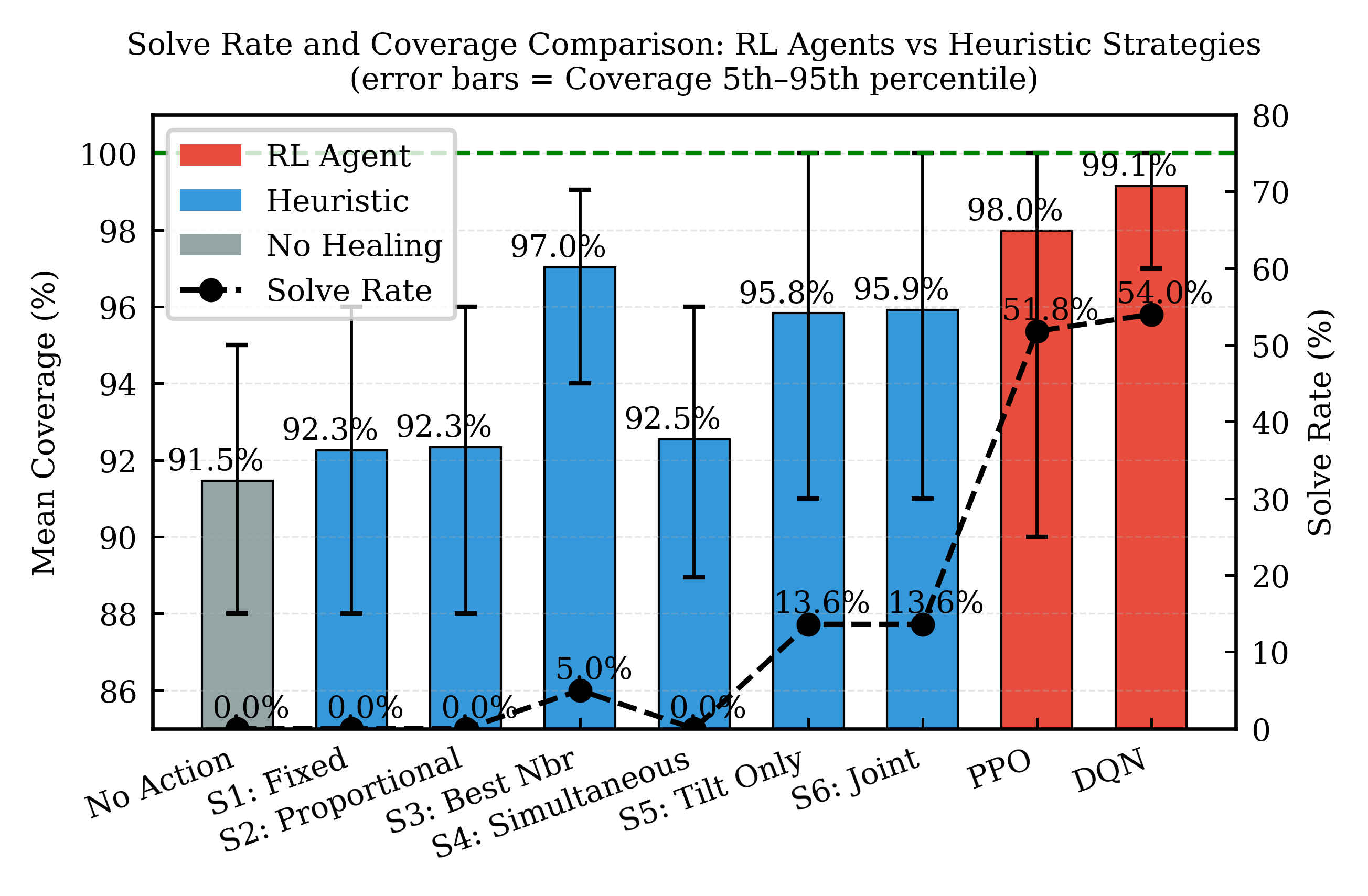}
	\caption{Mean coverage (left axis, error bars =
		5th--95th percentile) and solve rate (right axis)
		across all strategies, 500 test episodes each.}
	\label{fig:coverage}
\end{figure}

\begin{figure}[!t]
	\centering
	\includegraphics[width=\linewidth]{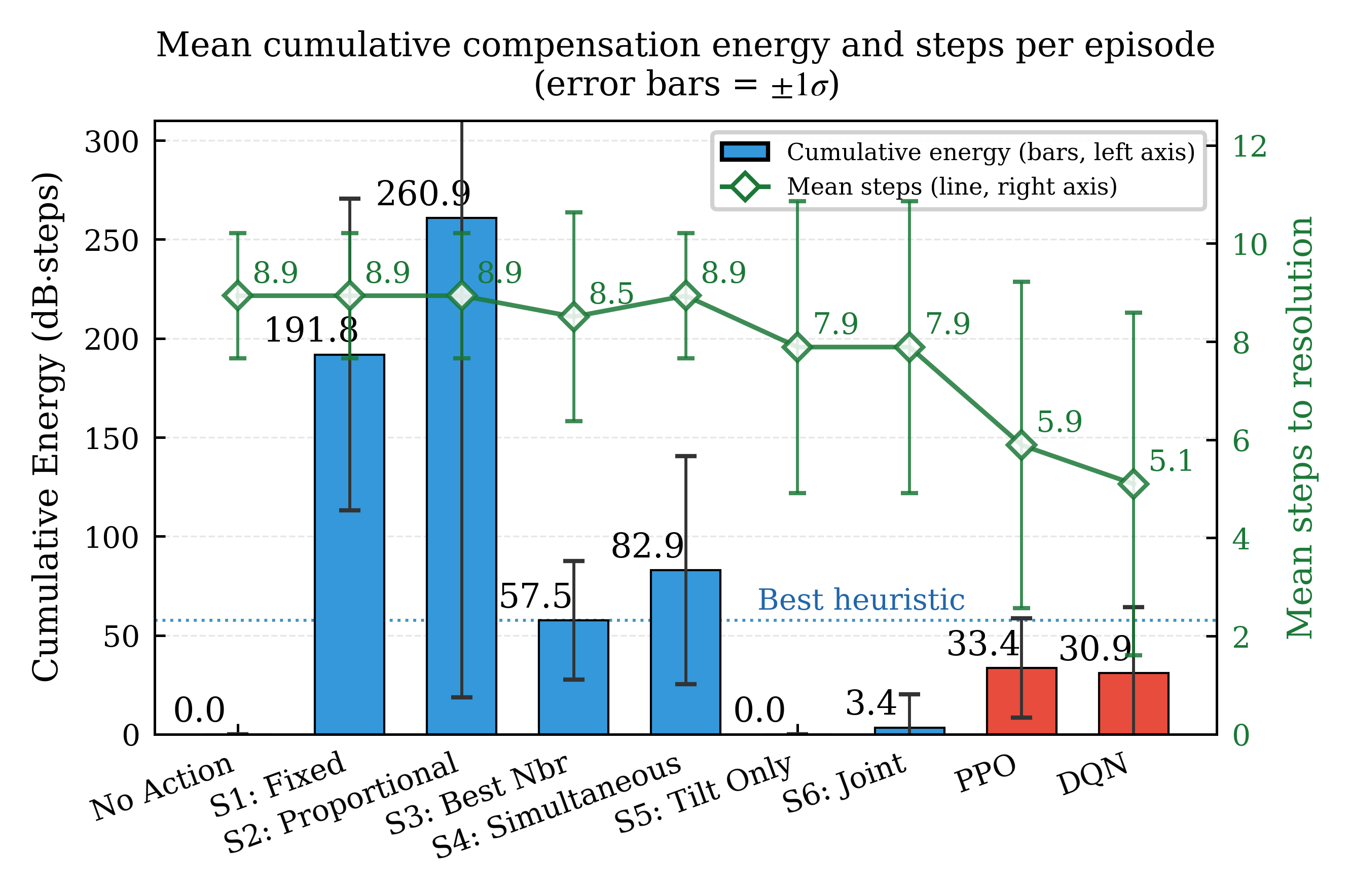}
	\caption{Mean cumulative compensation energy
		(left axis) and mean steps to resolution
		(right axis), error bars $=\pm1\sigma$.}
	\label{fig:energy}
\end{figure}

\begin{figure}[!t]
	\centering
	\includegraphics[width=\linewidth]{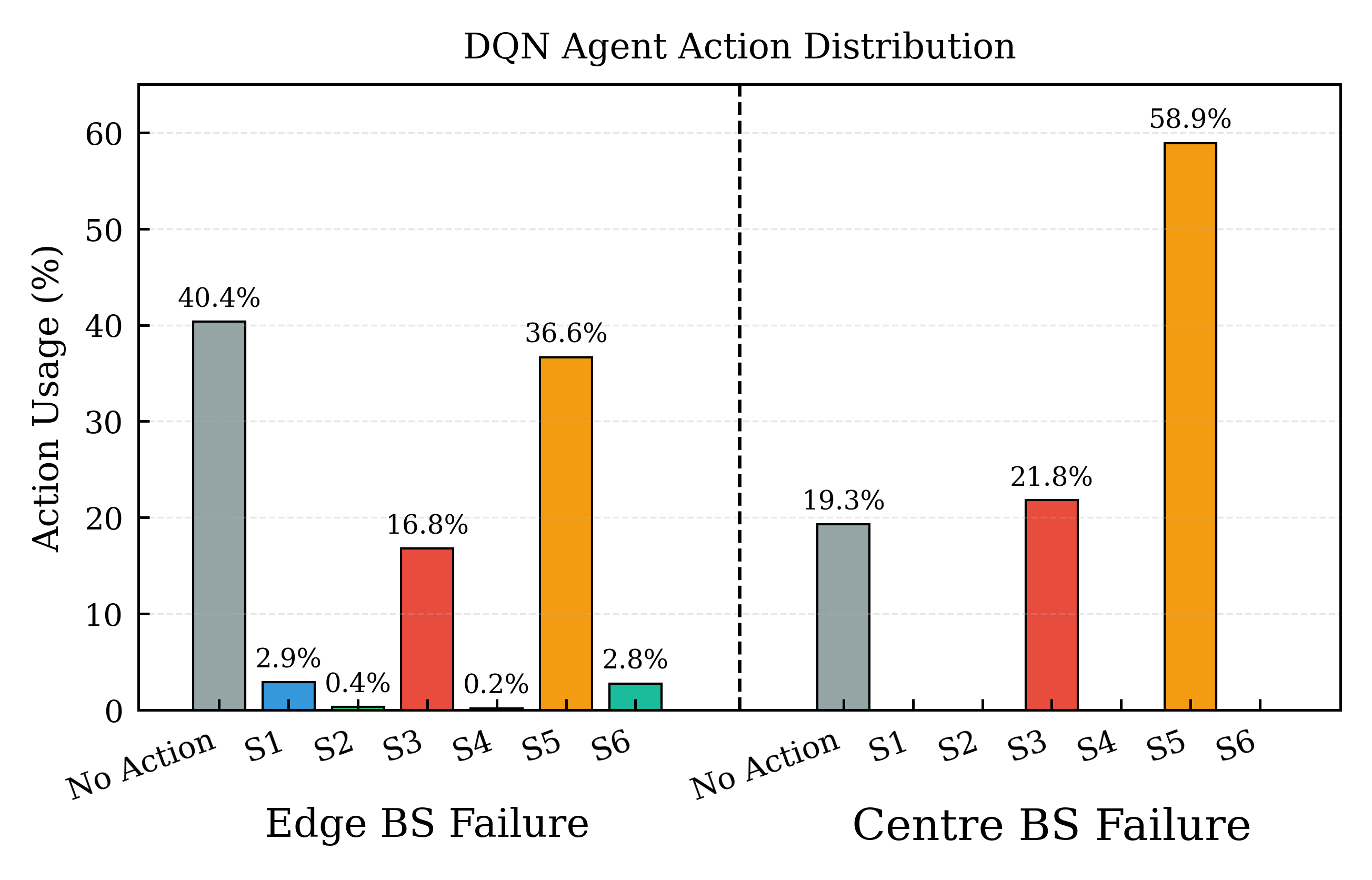}
	\caption{DQN action distribution for edge-BS
		(left) versus centre-BS (right) outage.}
	\label{fig:actions}
\end{figure}

The trained model is evaluated over $500$ test episodes using random seeds unseen during training, each initialising a fresh network and a randomly selected BS failure and is compared against the heuristic power-and-tilt strategies and PPO. The Coverage ratio, Solve rate, and Cumulative compensation energy are used as evaluation metrics. The \textit{Coverage ratio} ($\eta$) is defined as the fraction of UEs with SINR above $\gamma_{\rm th}$ at episode end. The \textit{Solve rate} is the fraction of episodes achieving complete restoration ($\eta{=}1.0 ~(100\%)$ corresponds to all outaged-UEs being rescued after the compensation action) within the horizon $T$, measuring full resolution rather than partial improvement. The \textit{Cumulative compensation energy} is defined as the episode-summed power boost $\sum_{t=0}^{T} \sum_{b} \Delta P_b^{(t)}$ applied across all neighbouring BSs at every step (units dB$\cdot$steps); it jointly captures both the per-step power magnitude and the number of steps required, so a strategy that resolves an outage quickly with modest boosts is rewarded over the one that is either slow or aggressive. A purely tilt-based action contributes zero energy by this definition, since no transmit power is altered. Figures~\ref{fig:coverage}, \ref{fig:energy}, \ref{fig:actions} show the mean values of the evaluation metrics over $500$ episodes and are summarized in Table~\ref{tab:results}.

\begin{table}[!t]
	\caption{Mean Evaluation Results over 500 Test Episodes}
	\label{tab:results}
	\centering
	\resizebox{\columnwidth}{!}{%
	\begin{tabular}{lcccc}
		\toprule
		\textbf{Strategy}
		& \textbf{Coverage}
		& \textbf{Solve}
		& \textbf{Energy}
		& \textbf{Steps} \\
		& \textbf{$\eta$\,(\%)}
		& \textbf{Rate, (\%)}
		& \textbf{(dB$\cdot$st.)}
		& \textbf{to Sol.} \\
		\midrule
		No Healing         & 91.5 &  0.0 &   0.0 & 8.9 \\
		S1: Fixed          & 92.3 &  0.0 & 191.8 & 8.9 \\
		S2: Proportional   & 92.3 &  0.0 & 260.9 & 8.9 \\
		S3: Best Nbr       & 97.0 &  5.0 &  57.5 & 8.5 \\
		S4: Simultaneous   & 92.5 &  0.0 &  82.9 & 8.9 \\
		S5: Tilt Only      & 95.8 & 13.6 &   0.0 & 7.9 \\
		S6: Joint P$+$T    & 95.9 & 13.6 &   3.4 & 7.9 \\
		PPO Agent          & 98.0 & 51.8 &  33.4 & 5.9 \\
		\textbf{DQN Agent} & \textbf{99.1}
		& \textbf{54.0} & \textbf{30.9} & \textbf{5.1} \\
		\bottomrule
	\end{tabular}}
\end{table}

It can be seen that the power-only heuristics S1, S2, and S4 achieve only 92.3--92.5\% coverage and zero solve rate despite consuming 82.9--260.9~dB$\cdot$steps of compensation energy, confirming the symmetric interference trap that boosting multiple equidistant neighbours raises interference as fast as signal, particularly under centre-BS outage. Tilt-based S5 and S6 achieve comparable coverage (95.8--95.9\%) and solve rate (13.6\%) at near-zero energy cost (0.0 and 3.4~dB$\cdot$steps respectively), confirming that antenna tilt extends coverage directionally without raising the interference floor.
S3 (best single neighbour) is the strongest heuristic at 97.0\% mean coverage by concentrating the power budget on one well-positioned neighbour,
avoiding the multi-neighbour interference penalty. 

The DQN agent outperforms every heuristic, reaching 99.1\% mean coverage and a 54\% solve rate, an 11$\times$ improvement in solve rate over the best power heuristic (S3, 5.0\%) and 4$\times$ improvement over tilt heuristic---while using only 30.9~dB$\cdot$steps of mean cumulative energy, less than S1, S2, S3, and S4, and resolving outages in 5.1 steps on average versus 7.9--8.9 for all heuristics. The DQN agent therefore simultaneously dominates on coverage, solve rate, \emph{and} compensation
cost, a combination no single heuristic achieves. PPO trails DQN on every metric (98.0\% coverage, 51.8\% solve rate, and 33.4 dB$\cdot$steps of compensation energy) but still exceeds all heuristics, consistent with off-policy experience replay DQN offering an advantage over on-policy PPO learning in this discrete action space.

Fig.~\ref{fig:actions} shows the DQN action distribution split by failure geometry. For edge-BS outage, the agent uses S3 (16.8\%) and S5 (36.6\%) as its primary active strategies, with the remaining power-only actions (S1, S2, S4) combining for under 4\%. For centre-BS outage, the learned policy concentrates entirely on three actions---no-action (19.3\%), S3 (21.8\%), and S5 (58.9\%)---assigning effectively zero weight to S1, S2, S4, and S6. The agent learns to avoid power-only and joint actions specifically under centre-BS failure, where they cannot rescue UEs without raising interference, and instead relies almost exclusively on tilt.
This action-pruning behaviour is learned purely from the reward signal with no explicit geometric input.

Under edge-BS outage, 40.4\% of decisions correspond to $A_0$ (no action), making it the most frequently selected action. Since the reward function in (\ref{eq:reward}) penalizes collateral outage, the agent learns to withhold intervention when further compensation risks disrupting already-served UEs. Similarly, under centre-BS outage, the agent prefers inaction over power-based compensation, having learned that power adjustments are ineffective under symmetric interference conditions.

\section{Conclusion}
\label{sec:conclusion}

This letter presented an end-to-end SON self-healing framework that integrates three-class outage detection with DRL-based outage compensation. Results demonstrate that jointly optimizing transmit power and antenna tilt enables substantially better recovery than conventional power- or tilt-based heuristics while requiring lower compensation energy and fewer recovery steps. Analysis of the learned policy reveals geometry-aware behavior: the agent suppresses ineffective power actions under centre-BS outage and preferentially exploits tilt adjustments to avoid the symmetric interference trap. Future work will extend the framework to multi-cell
simultaneous outages.

\bibliographystyle{IEEEtran}

\end{document}